\documentclass[mathleft
]{an}
\usepackage{graphicx}
\usepackage{times}
\def\teff{$T_{\rm eff}$}
\def\logg{$\log g$}
\def\vsini{$V\!\sin i$}
\def\kms{km~s$^{-1}$}
\def\cd  {{$\mbox{c~d}^{-1}$}}

\begin{document}

\Pagespan{789}{}
\Yearpublication{2006}%
\Yearsubmission{2005}%
\Month{11}%
\Volume{999}%
\Issue{88}%

\title{Amplitude variations of the CoRoT
Be star 102719279}

\author{J. Guti\'errez-Soto\inst{1,2}\fnmsep\thanks{Corresponding author:
  \email{jgs@iaa.es}\newline}
\and T. Semaan\inst{2}
\and  R. Garrido\inst{1}
\and F. Baudin\inst{3}
\and A.-M. Hubert\inst{2} 
\and C. Neiner\inst{2}
}
\titlerunning{Amplitude variations of the CoRoT
Be star 102719279}
\authorrunning{J. Guti\'errez-Soto et al.}
\institute{Instituto de Astrof\'{\i}sica de Andaluc\'{\i}a (CSIC), Apartado 3004, 18080, Granada, Spain
\and 
GEPI, Observatoire de Paris, CNRS, Universit\'e Paris Diderot; place Jules Janssen 92195 Meudon Cedex, France
\and
Institut d’Astrophysique Spatiale (IAS), b\^{a}timent 121, 91405 Orsay Cedex, France
}

\received{}
\accepted{}
\publonline{later}


\abstract{%
  The Be star 102719279 is an interesting target observed with CoRoT during two runs, giving us the possibility to study the stability of
the detected frequencies and to search for any correlation with outburts, as it was recently found in the Be star HD 49330 (Huat et al. 2009).
The light curve of the star 102719279 shows fadings, multiperiodicity, stable and transient frequencies, etc.
The short-term variations of the light curve are probably produced by non-radial pulsations together with some material that is ejected from
the star, which produces the transient frequency. It is should be noted that the two main frequencies are synchronized and have the maximum
amplitude just before the outburst.
}

\maketitle

\section{Introduction}

Be stars are main-sequence or slightly evolved B stars surrounded by
an equatorially condensed disk fed by discrete mass-loss events, also called outbursts. Although Be stars are rapid rotators, their rotational
velocity does not reach the break-up velocity (they rotate in average at about 90\%, Fr\'emat et al. 2005) and the disk formation mechanisms 
are still unknown (Porter \& Rivinius 2003).
Be stars are also known to show short-term variations caused by non-radial pulsations (eg. Diago et al. 2009, Neiner et al. 2009) and/or rotational modulation (eg. Balona 1990; Balona 2009).
Non-radial pulsations could provide the additional amount of angular momentum to eject the material to form the disk (Osaki 1986).
Based on an exhaustive spectroscopic campaign, Rivinius et al. (1998) showed that multimode pulsation was playing a triggering role in the mass transfer in the Be star $\mu$ Cen. 
In the star 102719279, thanks to the high-precision and long-duration photometric data of CoRoT, Guti\'errez-Soto et al (2008) found a clear correlation between the amplitude variations and the outburts. More recently, Huat et al. (2009) showed that the outburst occurred just after the moment the two main frequencies had the maximum amplitude in the Be star HD 49330.

The objective of this paper is to give more clues to understand the link between non-radial pulsations and the outbursts in Be stars by 
performing a detailed analysis of the amplitude variations of the light curve of the star 102719279.

\section{Observations}
The star 102719279 was observed in the exoplanetary field of CoRoT from 2007 February 3 to April 1 in the initial run (IR,
57 days) and from 2007 October 23 to 2008 Mars 3 in the first long run towards the Galactic anticenter
direction (LRA1, 131 days).

A low-resolution spectra (3250\--8800 \AA, R$\sim$400) was obtained in 2008, Mars 11 with CAFOS at the 2.2m telescope in Calar Alto. 
In addition, in 2009, January 13 and 15, two medium-resolution spectra in the blue and red bands were acquired with VLT-FLAMES in medusa mode (R$\sim$7000, ESO programme 082.D-0839(A), PI: Coralie Neiner). See Semaan et al. (2010) in these proceedings for a detailed explanation of the observations and reduction.

\begin{figure*}
\centering
\includegraphics[width=12cm,clip]{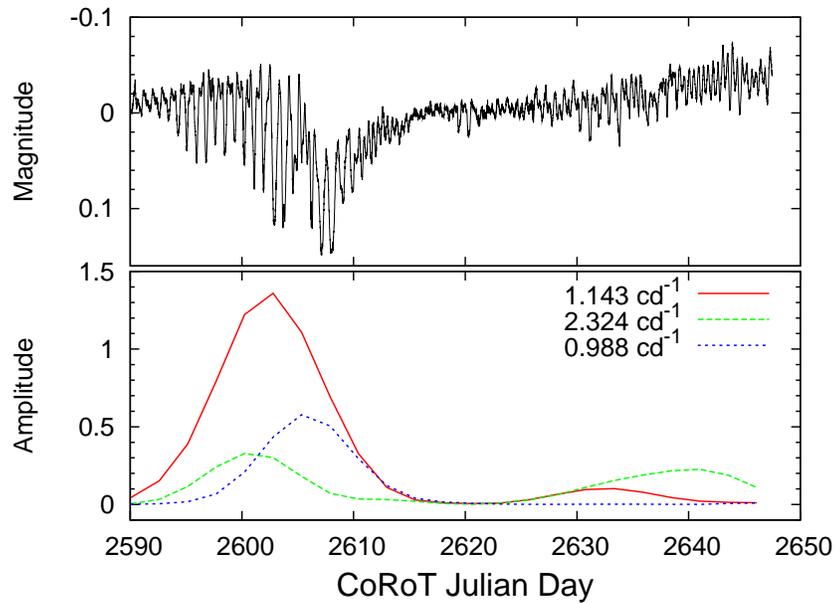}
\caption{In upper panel, light curve of the IR dataset and, in bottom panel, amplitude as a function of time (in arbitrary units) of the frequencies with largest amplitudes.}
\label{fig:AMP:IR}
\end{figure*}

\begin{figure*}
\centering
\includegraphics[width=12cm]{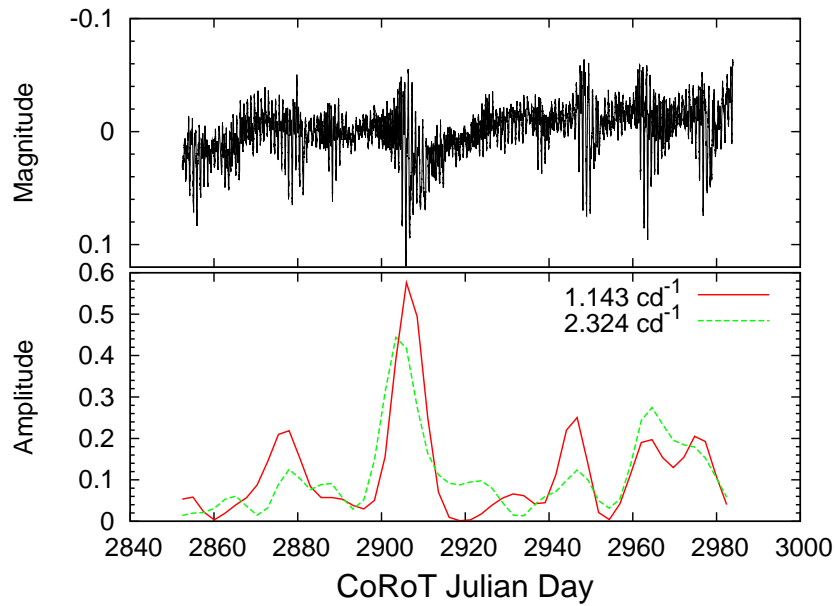}
\caption{Idem as Fig.~\ref{fig:AMP:IR}, but for the LRA1 dataset.}
\label{fig:AMP:LRA1}
\end{figure*} 

\section{The star 102719279}

The star 102719279 is an early B2.5e star with shell features visible in the Balmer and metal lines.  
The H$\alpha$ profile exhibits a double asymmetrical emission line with a deep and sharp absorption at the center. H$\alpha$ was observed in emission (I/I$_{c}$=4.5) in both spectra, in 2008 Mars 11 and in 2009 January 13-15. 

For the determination of the stellar parameters, we used the GIRFIT program (Fr\'emat et al. 2005). This method adjusts by least-squares fitting the observations with synthetic spectra interpolated in a grid of stellar fluxes computed with SYNSPEC (Hubeny \& Lanz 1995) and from model atmospheres calculated with TLUSTY (Hubeny \& Lanz 1995) and ATLAS9 (Kurucz, 1993). The best fit was found with the following \textit{apparent} parameters: a \teff~of $20000\pm1500$K, \logg~of $3.5\pm0.2$ dex and a \vsini~of $270\pm20$ \kms. We only used the photospheric lines without shell contamination. The parameters were not corrected from veiling caused by the emission nor from the effects of the stellar flattening and gravitational darkening caused by the rapid rotation (Fr\'emat et al. 2005).

From the fundamental parameters we see that the star is located inside both $\beta$ Cep and SPB instability strips.
In addition, we derived a rotational frequency of $1.19 \pm 0.3$ \cd\ from the radius calibration of Balona (1995). 
The star is seen equator-on (i $\sim 90$ deg).

\section{Analysis of the CoRoT light curve}

The light curves of the IR and the LRA1 (upper panel of Figs.~\ref{fig:AMP:IR} and~\ref{fig:AMP:LRA1} respectively) clearly show short-term variability. 
We also observe some sudden changes with amplitudes of several hundredths of magnitude, also called outbursts.
In this star, due to the high inclination of the line of sight, the ejected material produces an obscuration of part of the photosphere and thus we see a fading.

\begin{figure}
\includegraphics[width=8cm]{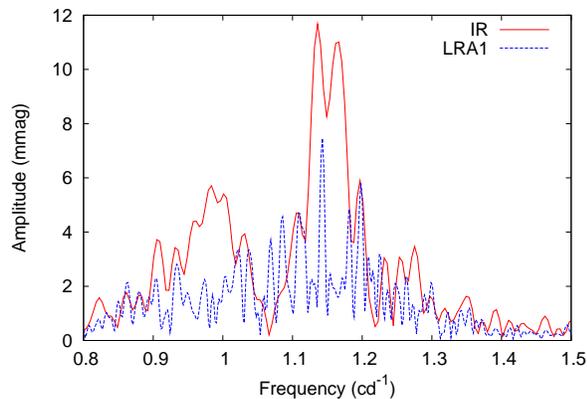}
\caption{Periodograms of the IR (red) and the LRA1 (blue) light
curves in the 0.8\--1.5 \cd\ frequency range.}
\label{fig:DFT:1.14}
\end{figure}

\begin{figure}
\includegraphics[width=8cm]{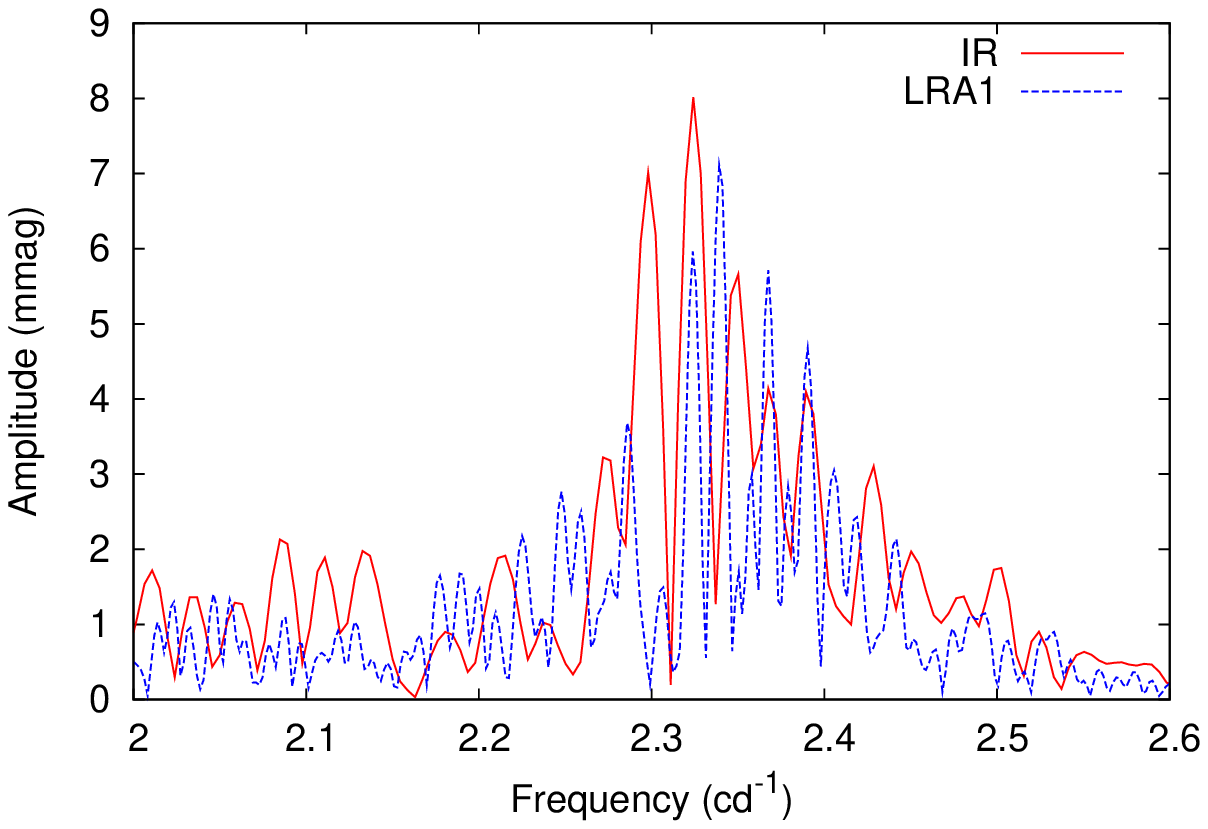}
\caption{Periodograms of the IR (red) and the LRA1 (blue) light
curves in the 2\--2.6 \cd\ frequency range.}
\label{fig:DFT:2.32}
\end{figure}

In order to study the short-term variations, we perform a standard Fourier analysis of both light curves. 
Multiple significant frequencies are detected in the periodograms in the 0.8\--1.5 and 2\--2.6 \cd\ frequency ranges. 
In Figs.~\ref{fig:DFT:1.14} and~\ref{fig:DFT:2.32} we display the periodogram of both datasets (IR in solid red and LRA1 in blue dashed line) between 0.8\--1.5 \cd\ 
and 2\--2.6 \cd\ respectively.
We also find more significant peaks in other frequency ranges, but they are probably combinations or harmonic of the frequencies with the largest amplitudes.

From a first look at the periodograms we see that the amplitudes of the detected frequencies change substantially with time. 
Several frequencies seem to appear and dissapear from the IR to the LRA1. 
However, other frequencies are detected in both runs, as for example 2.324 \cd, although with different amplitude.
We have summarised in Table~\ref{table:freq} the 5 frequencies with the largest amplitude detected separately in each dataset.
Note that, in the LRA1, for which we have better resolution, there are no frequencies that are exactly the double of another.
It is also important to highlight that the sinusoidal fittings with these 5 frequencies are not accurate enough to reproduce the light curves, and many more frequencies are needed, specially close to the times of the fadings. In addition, a frequency at 0.988 \cd\ is only detected in the IR. 

In order to determine the variations of the amplitudes of the main frequencies as a function of time, we 
have employed a method based on a wavelet analysis, using
the well known Morlet wavelet. This method was described and applied
to solar seismic data in Baudin et al. (1994) and to a CoRoT light curve of the Be star HD49330 in Huat et al. (2009). 
The frequency resolution (the full width at half maximum of the
wavelet in the Fourier domain) used in the present case is 1 μHz,
corresponding to a time resolution of ∼2.5 days.

We display in the lower panel of Figs.~\ref{fig:AMP:IR} and~\ref{fig:AMP:LRA1} the amplitude (in arbitrary units) of 
the frequencies 1.143, 2.324 and 0.988 \cd\ (the latter only detected in the IR).
The amplitudes of the main frequencies (1.143 and 2.324 \cd) 
seem to be synchronized reaching the maximum amplitude very
close to the time when the outbursts occur. This is seen in the fading of the IR and in most of the largest fadings of the LRA1.
The amplitudes are maximum just before the sudden decrease of the flux, as Huat et al. (2009) found for the Be star HD 49330.



In the lower panel of Fig.~\ref{fig:AMP:IR} we see that the frequency 0.98 \cd\ is only detected during the first 20 
days of the IR, reaching the maximum amplitude when the flux of the star starts to decrease. In addition, it occurs just after the time of the maximum amplitude of the main frequencies (1.143 and 2.324 \cd). Stefl et al. (1998) 
detected similar transient variations in other Be stars and linked
them to material in the inner or outer part of the circumstellar disk, which appeared at times of outbursts.
The transient period is normally 10\% of the genuine ones, which matches
very well with our values. 

\begin{table}
\centering
\footnotesize
\caption{Table with the 5 largest-amplitude frequencies for the IR and the LRA1 datasets.}
\label{table:freq}
\begin{tabular}{ccc|ccc}\hline
\multicolumn{3}{c|}{IR} & \multicolumn{3}{c}{LRA1} \\
\hline
\multicolumn{2}{c}{Frequency}   &Amplitude & \multicolumn{2}{c}{Frequency}&Amplitude\\ 
(\cd) & ($\mu$Hz)& (mmag) & (\cd) & ($\mu$Hz)& (mmag)\\
\hline
1.133 & 13.113& 11.4&1.143 & 13.227& 7.3\\
1.168 &	13.519& 11.6&2.339 & 27.074& 6.9\\ 
1.156 &	13.380& 9.4&2.324  & 26.898& 5.9\\
2.324 &	26.898& 8.0&1.198  & 13.866& 5.5\\
0.988 &	11.435& 5.8&2.368  & 27.407& 5.3\\
\hline
\end{tabular}
\end{table}

\section{Conclusions}

The light curve of the Be star 102719279 shows short-term multiperiodic variability. In addition, the amplitudes of the frequencies change substantially between and during the two studied datasets. The variability could be interpreted as non-radial pulsations and the changes in the amplitudes as the beating of the modes. We would need pulsating models that take into account the rapid rotation to confirm this interpretation.

In addition, the light curve of this star shows fadings or ejections of material that hides part of the photosphere.
In Guti\'errez-Soto et al. (2008) we performed a preliminary analysis of the first dataset (IR) and we found a correlation between the amplitude of the main frequencies and the fadings. Here we have studied in more detailed the amplitude variations of the same dataset (IR) and of a new one (LRA1), which contains more fadings. We have found in both datasets that the ejection of matter occurs just after the time when the amplitudes of the main frequencies are synchronised and are at the maximum value, as seen in Huat et al. (2009) for the Be star HD 49330.

\acknowledgements
The CoRoT space mission, launched on December 27th
2006, has been developed and is operated by CNES, with the contribution
of Austria, Belgium, Brasil, ESA, Germany and Spain. We wish to thank the
CoRoT team for the acquisition and reduction of the CoRoT data.
The FEROS data are being obtained as part of the ESO programme 082.D-0839 (PI: Neiner).
The work of JGS is supported by a JAE-DOC contract by the CSIC.

\newpage


\begin{thebibliography}{}
 
  \bibitem{} Balona, L. 1995, MNRAS, 277, 1547
  \bibitem{} Balona, L.. 2009, Stellar Pulsation: Challenges for theory and observation: Proceedings of the International Conference. AIP Conference Proceedings, Volume 1170, pp. 339-350 
  \bibitem{} Baudin, F., Gabriel, A., \& Gibert, D. 1994, A\&A, 285, L29
  \bibitem{} Diago, P.D., Guti\'errez-Soto, J., Auvergne, M., et al. 2009, A\&A, 506, 125
  \bibitem{} Fr\'emat, Y, Zorec, J, Hubert, A.-M. \& Floquet, M. 2005, A\&A, 440, 305
  \bibitem{} Guti\'errez-Soto, J., Neiner, C., Hubert, A.-M. et al. 2008, CoAst, 157, 70
  \bibitem{} Huat, A.-L., Hubert, A.-M., Baudin, F. et al. 2009, A\&A, 506, 95
  \bibitem{} Hubeny, I. \& Lanz, T. 1995, ApJ, 439, 875
  \bibitem{} Kurucz, R. L. 1993, CD-ROM No. 13 (Cambridge, Mass.: Smithsonian
Astrophysical Observatory)
  \bibitem{} Neiner, C., Guti\'errez-Soto, J., Baudin, F. et al. 2009, A\&A, 506, 143-151 
\bibitem{} Osaki, Y. 1986, PASP, 98, 30
  \bibitem{} Porter, J., Rivinius, Th. 2003, PASP, 115, 1153 
   \bibitem{} Rivinius et al. 1998, in Proc. ESO Workshop on Cyclical Variability in
Stellar Winds, ed. L. Kaper \& A. W. Fullerton (Berlin: Springer),
207
  \bibitem{} Stefl et al., 1998, MONS workshop
\end{thebibliography}
\end{document}